\documentclass[submission,copyright,creativecommons]{eptcs}
\providecommand{\event}{HAS 2014} 
\usepackage{amsmath,amssymb}
\usepackage{amsmath,amsthm,amssymb,graphicx,float}
\usepackage{algorithm, algpseudocode}
\usepackage[titletoc,title]{appendix}
\usepackage{hyperref}
\DeclareMathOperator*{\argmin}{arg\,min}
\DeclareMathOperator*{\cl}{cl}
\newtheorem{theorem}{Theorem}

\newtheorem{definition}[theorem]{Definition}

\newtheorem{proposition}[theorem]{Proposition}

\title{Safe Neighborhood Computation\\ for Hybrid System Verification}
\author{Yi Deng\thanks{YD and AAJ would like to acknowledge the support of NSF CAREER grant CNS-0953976.}
\institute{ECSE Department\\
Rensselaer Polytechnic Institute}
\email{dengy3@rpi.edu}
\and
A. Agung Julius
\institute{ECSE Department\\
Rensselaer Polytechnic Institute}
\email{juliua2@rpi.edu}
}
\def\titlerunning{Safe Neighborhood Computation for Hybrid System Verification}
\def\authorrunning{Yi Deng \& A. Agung Julius}
\begin{document}
\maketitle

\begin{abstract}
For the design and implementation of engineering systems, performing model-based analysis can disclose potential safety issues at an early stage. The analysis of hybrid system models is 
in general difficult due to the intrinsic complexity of hybrid dynamics. In this paper, a simulation-based approach to formal verification of hybrid systems is presented.
\end{abstract}

\section{Introduction}
Hybrid systems exhibit both discrete and continuous dynamics. The system state can flow continuously, and can also jump by triggering an event (transition). As an important application in the research of hybrid systems, safety verification is concerned with whether a specified set of unsafe states can be reached by the system from the initial set. One direct approach is to compute or over-approximate the set of all reachable states~\cite{
Girard2006efficient,
Guernic2009,Kurzhanski2000,Varaiya2000}, and then check the intersection with the unsafe set. The verification problem has also been investigated by using the abstraction approach, i.e., to construct a system model with a smaller or even finite state space, whose language is equivalent to or includes that of the original system~\cite{Tabuada2009}. Performing analysis of the abstraction is relatively easy, and allows us to verify properties of the original system. Various effective methods for system abstraction have been proposed~\cite{Alur2002reachability,
DInnocenzo2007_Approximate,Girard2010}. 
Reachable set computation, system abstraction, and some other approaches such as barrier certificate construction~\cite{Prajna2004safety} are capable of formally proving the system safety; but formal verification often comes at the price of conservatism and limited scalability.  

As complementary verification methods, randomized approaches have been proposed to strategically explore the state space with tools such as Rapidly-Exploring Random Trees (RRTs) and Probabilistic RoadMaps (PRMs)~\cite{Bhatia2004,Branicky2005}. By simulating trajectories from the initial set, one can falsify the system safety, or evaluate probabilistic safety. The randomized approaches are easy to implement because they are simulation-based; but usually a large number of trajectories need to be simulated, and no formal verification can be achieved.

It is possible to bridge the simulation-based approach and formal verification~\cite{Donze2007,Julius2007}: with finitely many simulations run for the sampled initial states, one can verify the safety of not only the samples but also infinitely many candidates in the initial set with mathematically proved guarantee.
As in \cite{Julius2007}, a  tube surrounding each simulated trajectory is computed, which over-approximates the reachable set for a neighborhood of initial states around the simulated one.  
If the simulated trajectory is safe, any trajectory initiated from the neighborhood must be safe, and moreover, must trigger the same event sequence as the simulated trajectory does. Such neighborhood is called a robust neighborhood, which has both uniform safety and transition properties. If the initial set can be fully covered by the robust neighborhoods of finitely many simulated trajectories, then its transition and safety properties are verified. However, we will see in Section \ref{sec:safe} that for pure safety verification problems the applicability of the robust neighborhood approach is limited, since the computed robust neighborhood can vanish due to the transition property required rather than safe property. 

Motivated by the robust neighborhood approach, we propose an algorithm for safe neighborhood computation in the present work. As its name implies, all trajectories initiated from a safe neighborhood are guaranteed safe for certain time horizon, although their event sequences are possibly different from that of the simulated trajectory. The safe neighborhood computed for any initial state is essentially a superset of the robust neighborhood, and may have non-zero measure even if the robust neighborhood vanishes. Consequently, for some initial state that cannot be covered by any robust neighborhood, the computed safe neighborhood is able to cover it; for some initial set where the coverage following \cite{Julius2007} never reaches $100\%$, the present approach using safe neighborhoods is able to reach full coverage and verify complete safety.

\section{Safe Neighborhood Approach}
\label{sec:safe}
\subsection{Hybrid Automata Formulation}
A hybrid automaton is a tuple $H = (L\times X, L_0\times X_0, D, E,$ $Inv)$~\cite{Alur1995}.

The state space is $L\times X$, where $L$ denotes the sets of discrete states (also called locations) and $X$ denotes the set of continuous states. The initial set is $L_0\times X_0\subset L\times X$. 

Each location $\ell\in L$ is associated with an invariant set $Inv(\ell)\subset X$. If the system is at location $\ell$, the continuous state $x\in X$ must satisfy $x\in Inv(\ell)$. The system dynamics $D$ maps a pair $(\ell,x)$ to $\dot{x}$, the time derivative of $x$. Let $D^\ell$ denote the restriction of $D$ to $\{\ell\}\times X$. At location $\ell$, the system state evolves continuously according to $D^\ell$ until an event (an instantaneous transition) $e:=(\ell,\ell',g,r), e\in E$ occurs. The event is guarded by $g\subset Inv(\ell)$. Namely, a necessary condition for the occurrence of $e$ is $x\in g$. After the event, the discrete state changes from the source $\ell$ to the target $\ell'$, and the continuous state is reset according to the reset map $r:Inv(\ell)\rightarrow Inv(\ell')$. Let $(\ell,x)$ denote the system state that triggers $e=(\ell, \ell', g, r)$. Then the reset state is $(\ell', r(x))$.

A trajectory $\rho(\ell_0, x_0)$ of the hybrid system is the solution of $(\ell,x)$ initiated from $(\ell_0,x_0)$. Clearly,  $\rho(\ell_0, x_0)$ is piece-wise continuous. At each location $\ell$, we write $\xi^\ell(t,x_0^\ell)\in Inv(\ell), t^\ell_0\le t\le t_{end}^\ell$ as the solution of $x$, where $x_0^\ell=\xi^\ell(t_0^\ell,x_0^\ell)$ is the initial condition in $\ell$, and for $t_0^\ell\le t\le t_{end}^\ell$ the function $\xi^\ell$ satisfies the differential equation $\frac{\partial\xi^\ell(t,x_0^\ell)}{\partial t}=D^\ell(\xi^\ell(t,x_0^\ell))$.

Consider the system state that reaches the boundary of the invariant set at the time instant $t^\ell_{end}$, i.e., $\xi^{\ell}(t^\ell_{end},x_0^\ell)\in\partial Inv(\ell)$. If there exits $\tau>0$ such that for all $\tau_1\in(0,\tau)$, $\xi^{\ell}(t^\ell_{end}+\tau_1,x_0^\ell)\not\in Inv(\ell)$, then we say the continuous state is evolving outward $Inv(\ell)$ at the boundary.

Let $\partial Inv(\ell)_{out}$ denote part of the boundary $\partial Inv(\ell)$ where the continuous state is evolving outward $Inv(\ell)$, $G^\ell$ denote the set of guards such that the corresponding events all have $\ell$ as the source location. We assume for all $\ell$:
\vspace{-2pt}
\begin{enumerate}
\item For all $g_1, g_2 \in G^{\ell}$, $g_1, g_2 $ are disjoint.
\item An event is forced to occur whenever $x\in\partial Inv(\ell)_{out}$. Without this assumption, the system state will get stuck at $\partial Inv(\ell)_{out}$, since it is not allowed to evolve outside $Inv(\ell)$. In addition, assume events can only be triggered at $\partial Inv(\ell)_{out}$. Define the active guards $G_{act}^\ell:=\{g\cap \partial Inv(\ell)_{out}\vert g\in G^\ell\}$.
\item  $\dot{x}=D^\ell(x)$ admits an unique global solution.
\item All the reset maps are continuous.
\end{enumerate} 

\subsection{Trajectory Robustness}
\begin{figure}[ht]
\begin{minipage}[b]{0.45\linewidth}
\centering
\setlength{\belowcaptionskip}{-10pt}
\includegraphics[width=\textwidth]{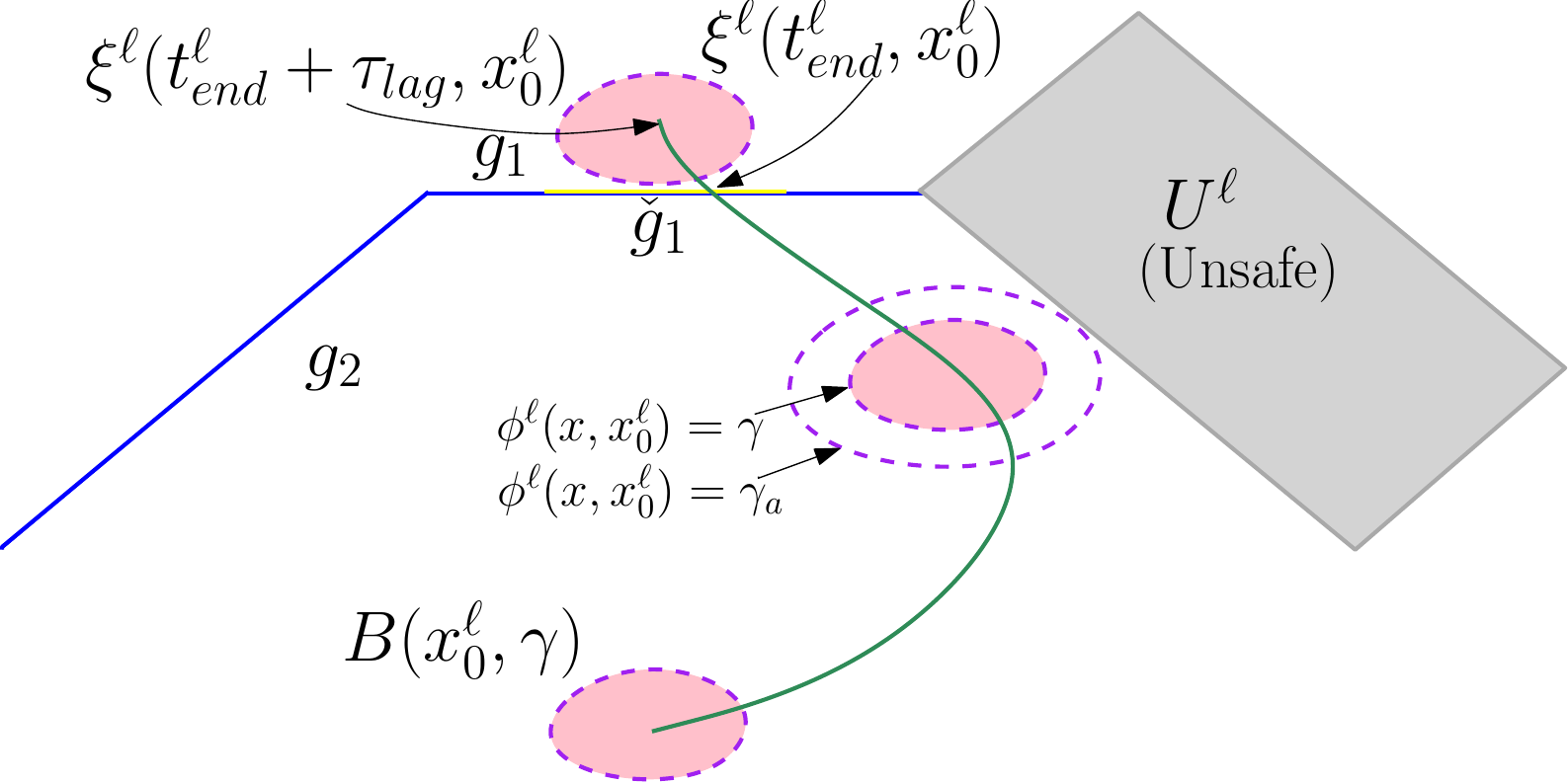}
\caption{Robust neighborhood computation.}
\label{fig_robust}
\end{minipage}
\hspace{0.5cm}
\begin{minipage}[b]{0.45\linewidth}
\centering
\setlength{\belowcaptionskip}{-10pt}
\includegraphics[width=\textwidth]{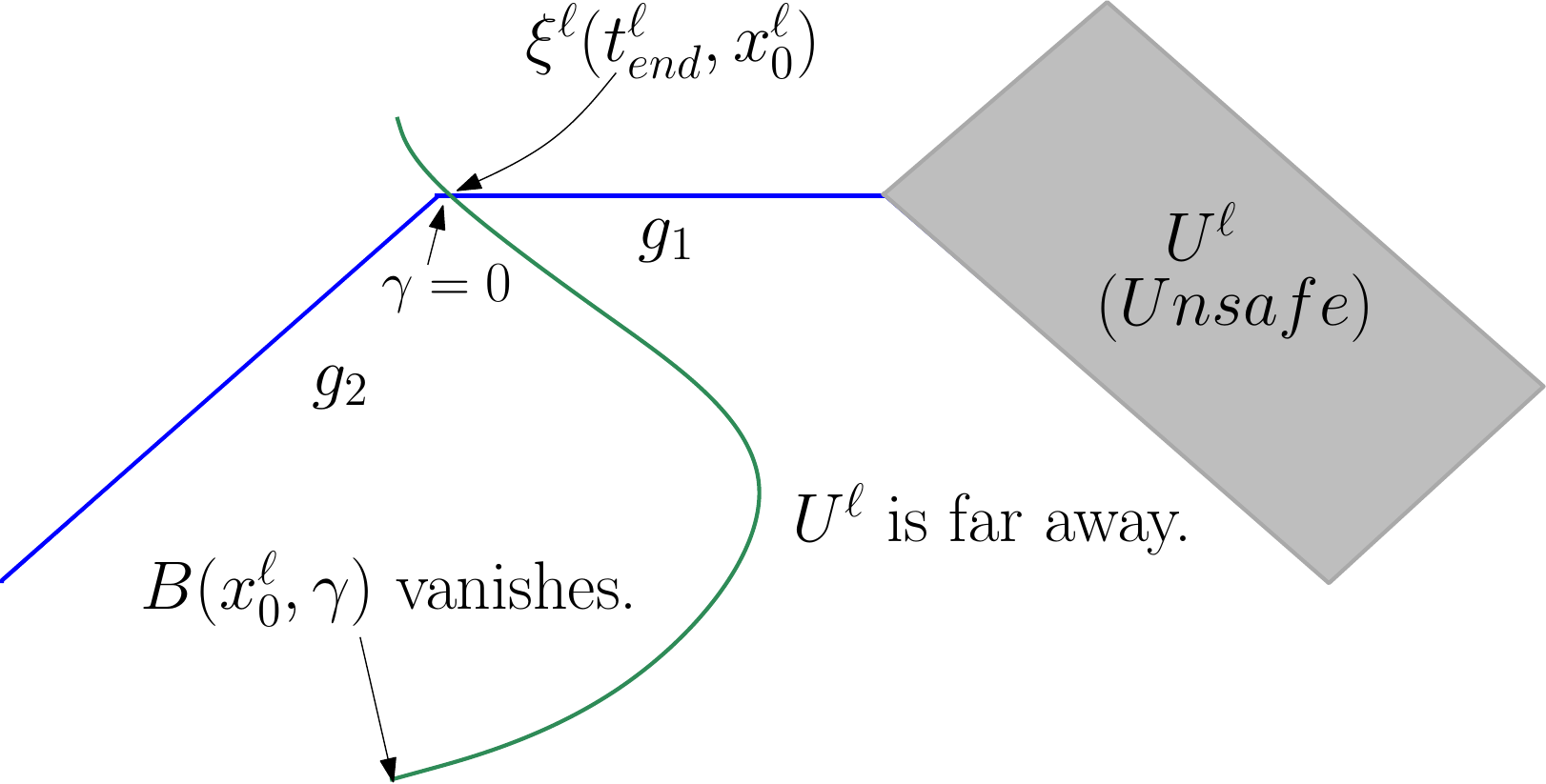}
\caption{Guard-critical trajectory.}
\label{fig_trivial}
\end{minipage}
\end{figure}

We briefly review the algorithm proposed in \cite{Julius2007}  for the computation of \emph{robust neighborhood} around a simulated initial state, which is based on the theory of bisimulation functions~\cite{Girard2007}.
\begin{definition}\emph{\cite{Girard2007}}
Let $\phi^\ell: X \times X \rightarrow \mathbb{R}$ be a pseudo-metric on the state space of the dynamical system $\dot{x} = D^\ell(x),x\in X$.
Let $\xi^\ell(t,x_{0}^\ell)$ denote the solution of $D^\ell$ under the initial condition $x_0$. If for any initial states $x_{0}^\ell$ and $\tilde{x}_{0}^\ell$, the function $\phi^\ell(\xi^\ell(t,x_{0}^\ell), \xi^\ell(t,\tilde{x}_{0}^\ell))$ is non-increasing with respect to time $t$, then $\phi^{\ell}$ is a bisimulation function between the system and itself.
\end{definition}
Consider a nominal trajectory $\rho(\ell,x_0^\ell)$ as shown in Fig. \ref{fig_robust}, which has been simulated for the time horizon of interest, $[t_0, t_{end}]$. The first segment of $\rho(\ell,x_0^\ell)$ is $\xi^\ell(t,x_0^\ell), t_0^\ell<t<t_{end}^\ell$, where $t_0^\ell=t_0$ is the initial time. At the time $t_{end}^\ell$, $\rho(\ell,x_0^\ell)$ leaves $\ell$ by triggering the event $e_1 = (\ell,\ell',g_1,r_1)$ , i.e., $\xi^\ell(t_{end}^\ell,x_0^\ell)\in g_1$. 
Define the \emph{avoided set} 
\begin{equation}
A^\ell:= U^\ell \cup (G_{act}^\ell \setminus \check{g}_1),
\label{eq:avd}
\end{equation} 
where $\check{g}_1$ is called the \emph{allowed part} of the guard $g_1$. We will formally define $\check{g}_1$ later.
Essentially, the robust neighborhood is to be computed based on the avoided set $A^\ell$,  so that all trajectories initiated from the robust neighborhood will not reach  $A^\ell$ in location $\ell$. 

Hence, the unsafe $U^\ell$ must be included in $A^\ell$, as well as the undesired part of guards $G_{act}^\ell \setminus \check{g}_1$. In this particular example shown in Fig. \ref{fig_robust}, the undesired part of guards $G_{act}^\ell \setminus \check{g}_1:= g_2\cup (g_1\setminus\check{g}_1)$, where $g_2$ is undesired because it triggers an event $e_2$ different from the event $e_1$  triggered by the nominal trajectory, while $\check{g}_1$ is excluded from $A^\ell$ since trajectories initiated from the robust neighborhood are allowed to reach $\check{g}_1$ and trigger $e_1$.
Because of the monotonicity of $\phi^\ell$, for any time $t>t_0^\ell$ and initial state $\tilde{x}_0^\ell$,
\begin{equation}
\phi^\ell(\xi^\ell(t,x_{0}^\ell), \xi^\ell(t,\tilde{x}_{0}^\ell)) \le \phi^\ell(\xi^\ell(t_0^\ell,x_{0}^\ell), \xi^\ell(t_0^\ell,\tilde{x}_{0}^\ell))
= \phi^\ell(x_{0}^\ell,\tilde{x}_{0}^\ell).
\end{equation}
Therefore, if $\tilde{x}_0^\ell$ satisfies 
\begin{equation}
\phi^\ell(x_{0}^\ell,\tilde{x}_{0}^\ell) < \gamma_a
:=\inf_{t\in [t_0^\ell,t_{end}^\ell]}\inf_{y \in A^\ell} \phi^\ell(\xi^\ell(t,x_{0}^\ell),y),
\label{eq:gamma}
\end{equation}
then for all $t\in [t_0^\ell,t_{end}^\ell]$, $\xi^\ell(t,\tilde{x}_{0}^\ell)\not\in A^\ell$.

The time horizon $[t_0^\ell,t_{end}^\ell]$ above may be too short, since $\rho(\ell,\tilde{x}_0^\ell)$ may leave $\ell$ later than $\rho(\ell,x_0^\ell)$ does. This time lag problem is handled by the $Shrinking$ procedure (proposed in \cite{Julius2007}, and can also be found in Algorithm \ref{alg:shrinking}): defined a preliminary robust neighborhood $B(x_0^\ell,\gamma_a):=\{\phi^\ell(x_0^\ell, \tilde{x}_0^\ell)<\gamma_a\}$,
and then shrinks $B(x_0^{\ell},\gamma_a)$ to a proper size $B(x_0^{\ell},\gamma)$ as the robust neighborhood. 
As a result, for some time lag $\tau_{lag}$ that does not exceed the specified parameter $\tau_{maxlag}$, all trajectories initiated from $B(x_0^{\ell},\gamma)$ are guaranteed to leave $Inv(\ell)$ before $t_{end}^\ell+\tau_{lag}$, and will not reach $A^\ell$ before they trigger $e_1$ at $\check{g}_1$. See Fig. \ref{fig_robust}. 

It is also proposed in \cite{Julius2007} how to compute the event time lead $\tau_{lead}$ such that all trajectories initiated from $B(x_0^{\ell},\gamma)$ are guaranteed to stay in $\ell$ before $t_{end}^\ell-\tau_{lead}$. We use $\tau_{maxlead}$ to denote an upper bound of the event time lead for the robust neighborhoods.

The allowed part of guard $\check{g}_1$ in Eq. \eqref{eq:avd} is defined according to the robust neighborhood computed for the next location reached by the nominal trajectory using similar steps as Eq. \eqref{eq:avd}, \eqref{eq:gamma}: let $B(x_0^{\ell'},\gamma')$ denote the robust neighborhood computed for the reset initial state $x_0^{\ell'}:=r_1(\xi^{\ell}(t^\ell_{end},x_0^\ell))$, then 
\begin{equation}
\check{g}_1:=r_1^{-1}(B(x_0^{\ell'},\gamma'))\cap g_1.
\label{eq:allow}
\end{equation}

Therefore, the robust neighborhood is computed in a recursive way, from the last location reached to the first location reached.

In the last location reached (denoted by $\mathfrak{l}$), the avoided set is defined in a form different form Eq. \eqref{eq:avd}.
\begin{equation}
A^\mathfrak{l} := U^\mathfrak{l}\cup G_{act}^\mathfrak{l}.
\end{equation}

Event time lag does not need to be considered, since $\mathfrak{l}$ is the last location reached. 

From the argument above, $B(x_0^{\ell},\gamma)$ has the following property:
\begin{proposition}
For all $\tilde{x}_0^\ell\in B(x_0^{\ell},\gamma)$, the trajectory $\rho(\ell,\tilde{x}_0^\ell)$ must trigger the same event sequence as the nominal trajectory $\rho(\ell, x_0^\ell)$ does. The time lead and lag for triggering the same event is bounded by $\tau_{maxlead}$ and $\tau_{maxlag}$ respectively. In all the locations reached except the last one, $\rho(\ell,\tilde{x}_0^\ell)$ must stay safe before it leaves the location. In the last reached location $\mathfrak{l}$,
$\rho(\ell,\tilde{x}_0^\ell)$ must stay safe for at least $[t_0^\mathfrak{l},t_{end}^\mathfrak{l}]$ as the nominal trajectory $\rho(\ell,x_0^\ell)$ does.
\end{proposition}
\subsection{Critical Trajectory}
Suppose in Fig. \ref{fig_robust}, the nominal trajectory reaches the closure of $g_2$, $g_1\setminus \check{g}_1$ or $U^\ell$, then clearly Eq. \eqref{eq:gamma} results in zero. Such a trajectory is called critical.
\begin{definition}[Critical Trajectory]
If a nominal trajectory reaches the closure of the avoided set in the robust neighborhood computation, then it is called a critical trajectory.
\end{definition}
Directly following from the algorithm in \cite{Julius2007}, the proposition below holds:
\begin{proposition}
The robust neighborhood computed for a nominal trajectory has zero measure if and only if the nominal trajectory is a critical trajectory.
\end{proposition}

Essentially, a critical trajectory has trivial robustness. There exists some infinitesimal perturbation of the trajectory that changes its transition or safety property.
In particular, we define \emph{guard-critical} trajectories, whose robust neighborhoods vanish due to guards rather than the unsafe set.
\begin{definition}[Guard-Critical Trajectory]
A critical trajectory that does not reach the closure of the unsafe set is called a guard-critical trajectory.
\end{definition}
Guard-critical trajectories can cause issues in safety verification problems, where only the safety property is of concern. As shown in Fig. \ref{fig_trivial}, the guard-critical trajectory triggers an event through $g_1$, but it also reaches the closure of $g_2$. By the robust neighborhood algorithm, the initial state $(\ell,x^\ell_0)$ cannot be covered by the robust neighborhood of any initial state. Consequently, if an initial set contains such $(\ell,x^\ell_0)$, it can never be covered fully by robust neighborhoods. On the other hand, the nominal trajectory $\rho(\ell,x^\ell_0)$ is far from unsafe. So the robust neighborhood approach does not work in a satisfactory way for the purpose of safety verification.

In this work, an adapted approach called safe neighborhood is proposed to deal with this issue. Essentially, for each nominal trajectory, the computed robust neighborhood has both uniform transition and safety properties, while the safe neighborhood has only uniform safety property. The latter is thus a superset of the former. 

\subsection{Safe Neighborhood Computation}
\begin{figure}[h]
\centering
\setlength{\belowcaptionskip}{-10pt}
\includegraphics[scale=0.8]{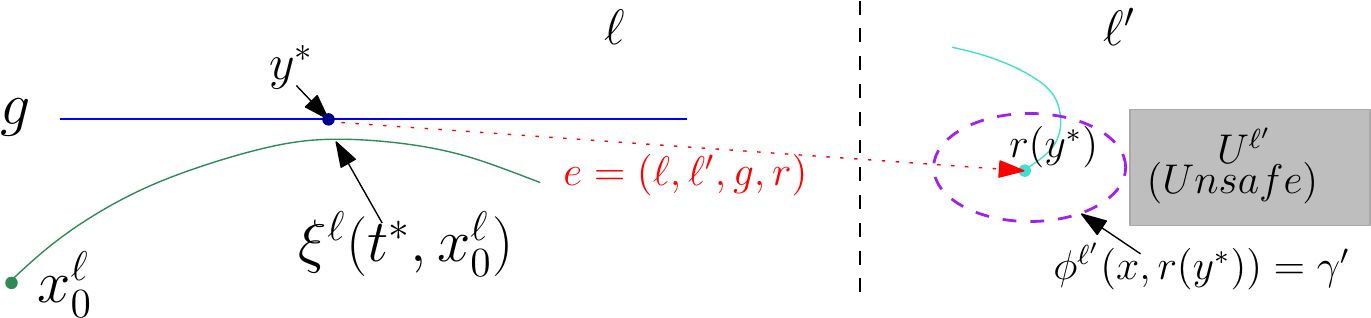}
\caption{Basic case of safe neighborhood computation.}
\label{fig_basic}
\end{figure}

\paragraph{Basic Case}
In order to illustrate the basic idea of safe neighborhood computation, first consider the simple case shown in Fig. \ref{fig_basic}. For simplicity, it is assumed the nominal trajectory $\rho(\ell,x_0^\ell)$ does not trigger any event; but it gets sufficiently close to the active part of guard $g_{act}:=g\cap\partial Inv(\ell)_{out}$ within the time horizon $[t_0^\ell,t_{end}^\ell]$ . The guard $g$ is associated with the event $e=(\ell,\ell',g,r)$. In the location $\ell'$, there are no guards. The unsafe set is assumed to be only in $\ell'$, i.e., $U^\ell$ is empty.  

\begin{algorithm}[H]
\begin{algorithmic}[1]
\caption{Basic case of safe neighborhood computation.}
\State compute
$(t^*,y^*)=\argmin\limits_{t\in[t_0^\ell,
t^\ell_{end}],y\in\cl(g_{act})}\phi^{\ell}(\xi^{\ell}(t,x_0^\ell), y)$\;
\Comment $\cl()$ gives the closure of a set.
\If{$\phi^{\ell}(\xi^{\ell}(t^*,x_0^\ell), y^*)\le d_{thr}$}
\State simulate a trajectory from $r(y^*)$ for the time horizon $t^*\le t\le t^\ell_{end}$\;
\State compute $\gamma'=\inf\limits_{y\in U^{\ell'}}\inf\limits_{t\in[t^*,t^\ell_{end}]}\phi^{\ell'}(\xi^{\ell'}(t,r(y^*)), y)$\;
\State define $\mathring{g}_{act} :=\{y\in g_{act}\vert \phi^{\ell'}(r(y), r(y^*))\ge \gamma'\}$
\State specify a time interval $\delta:=[t^*-\tau_{lead}, t^*+\tau_{lag}]$\;
\State compute $\gamma=\min\{\inf\limits_{t\in [t_0^\ell,t_{end}^\ell]\setminus\delta}\inf\limits_{y\in g_{act}}\phi^{\ell}(\xi^{\ell}(t,x_0^\ell),y), \inf\limits_{t\in\delta}
\inf\limits_{y\in\mathring{g}_{act}}\phi^{\ell}(\xi^{\ell}(t,x_0^\ell),y)\}$\;
\Else 
\State compute $\gamma=\inf\limits_{t\in [t_0^\ell,t_{end}^\ell]}\inf\limits_{y\in g_{act}}\phi^{\ell}(\xi^{\ell}(t,x_0^\ell),y)$\;
\EndIf
\State $Safe(x_0^\ell):=\{x|\phi^\ell(x,x_0^\ell)\le\gamma\}$
\label{alg:basic}
\end{algorithmic}
\end{algorithm}
\vspace{-10pt}

At the point $y^*$ and the time instant $t^*\in [t_0^\ell,t_{end}^\ell]$, the nominal trajectory and the guard $g$ get sufficiently close ($\phi^\ell$ attains its infimum, and the infimum is smaller than the specified threshold value $d_{thr}$, which corresponds to the first case in the if-else block of Algorithm \ref{alg:basic}). Since $U^\ell$ is assumed as empty, the bottleneck of robust neighborhood computation is in the guard. We simulate a \emph{branch trajectory} from $y^*$ for the rest of the time: $t^*\le t\le t^\ell_{end}$, which triggers $e=(\ell,\ell',g,r)$.
In the target location $\ell'$, there are no guards. We compute the infimum value $\gamma'$ of $\phi^{\ell'}$ generated by the branch trajectory and the unsafe set $U^{\ell'}$. Because of the monotonicity of $\phi^{\ell'}$, for all $t\in [t^*,t_{end}^\ell]$ and $x_0^{\ell'}\in\{x\vert\phi^{\ell'}(x,r(y^*))<\gamma'\}$, $\xi^{\ell'}(t,x_0^{\ell'})$ cannot reach $U^\ell$ (see arguments in the robust neighborhood computation). 

We thus define $\check{g}:=\{y\in g\vert \phi^{\ell'}(r(y), r(y^*))<\gamma'\}$ as the allowed part of $g$. For the specified time window $\delta:=[t^*-\tau_{lead}, t^*+\tau_{lag}]$,  consider $\mathring{g}_{act}:=g_{act}\setminus\check{g}$ as the avoided set; while for the reset of the time, $[t_0^\ell,t_{end}^\ell]\setminus\delta$, consider the entire $g_{act}$ as the avoided set. Specifically, we compute
\begin{equation}
\gamma=\min\{\inf\limits_{t\in [t_0^\ell,t_{end}^\ell]\setminus\delta}\inf\limits_{y\in g_{act}}\phi^{\ell}(\xi^{\ell}(t,x_0^\ell),y), \inf\limits_{t\in\delta}\inf\limits_{y\in\mathring{g}_{act}}\phi^{\ell}(\xi^{\ell}(t,x_0^\ell),y)\}.
\label{eq:min}
\end{equation}
Then for all $\tilde{x}_0^\ell\in Safe(x_0^\ell):=\{x|\phi^\ell(x,x_0^\ell)<\gamma\}$ and $t\ge t_0^\ell$, because of the monotonicity of $\phi^{\ell}$, 
\begin{equation}
\phi^\ell(\xi^\ell(t,\tilde{x}_0^\ell),\xi^\ell(t,x_0^\ell)) \le \phi^\ell(\xi^\ell(t_0^\ell,\tilde{x}_0^\ell),\xi^\ell(t_0^\ell,x_0^\ell))= \phi^\ell(\tilde{x}_0^\ell,x_0^\ell)<\gamma.
\end{equation}
As a result,  for all $ t\in [t_0^\ell,t_{end}^\ell]\setminus\delta$, $\xi^\ell(t,\tilde{x}_0^\ell) \not\in g_{act}$, while for all $t\in \delta$, $\xi^\ell(t,\tilde{x}_0^\ell)\not\in\mathring{g}_{act}$.
Namely, the trajectory $\rho(\ell,\tilde{x}_0^\ell)$ is allowed to escape from $\check{g}$ during $\delta$, and then stays in $\ell'$ safely for at least $t^\ell_{end}-t^*$. If no event has been triggered, $\rho(\ell,\tilde{x}_0^\ell)$ must stay in $\ell$ safely as the nominal trajectory  $\rho(\ell,x_0^\ell)$ does.
\vspace{-10pt}

\paragraph{General Case}
For more general cases, the safe neighborhood of a nominal trajectory $\rho(\ell_0,x_0)$ can be computed as in Algorithm \ref{alg:general}. The time horizon is $t_0\le t\le t_{end}$. For clarity, we denote the trajectory segments as $\{\xi^{\ell_i}(t,x_0^i),t_0^i\le t\le t_{end}^i\}_{i=1}^N$, where $N$ is the total number of events triggered.

The essential idea is as presented in the basic case: When the nominal trajectory gets sufficiently close to a guard, even if it does not actually trigger the corresponding event, we still simulate a branch trajectory according to the event. This is called a \emph{virtual event}. For the branch trajectory we compute the safe neighborhood. Part of guards that maps into the safe neighborhood of the branch trajectory is then considered as the allowed part. We exclude it from the avoided set for a short time window, and thus removed the bottleneck of the bisimulation function value. Clearly, the algorithm must be performed in recursive way.
The nominal trajectory can get sufficiently close to multiple guards in one location, and it can also get sufficiently close to guards in sequentially reached locations. For each location, not only the event triggered by nominal trajectory itself by also all the virtual events need to be considered. We call the collection of triggered events and virtual events the \emph{event tree} associated with the nominal trajectory.  
\begin{algorithm}
\caption{Safe neighborhood computation for a general trajectory.}
\label{alg:general}
\begin{algorithmic}[1]
\Procedure{SafeNeighborhood}{$\ell_0$, $x_0$, $t_0$, $t_{end}$}
\For{$i\gets N$ to $1$}
\State $d^u_{i}\gets\inf\limits_{t\in [t_0^i,t_{end}^i]}\inf\limits_{y\in U^{\ell_i}}\phi^{\ell_i}(\xi^{\ell_i}(t,x_0^i),y)$
\State $d_i\gets\min\{d^u_{i}, d_{thr}\}$
\State $\mathcal{T}_i\gets\{t\in [t_0^i,t_{end}^i]\vert ProximalGuards(\ell_i,x_0^i,t,d_i)\neq\emptyset\}$
\State $\mathcal{T}\gets\mathcal{T}_i$, $k\gets 0$, $d^{(k)}\gets\infty$
\Comment{$\mathcal{T}$ is the set of time instants when the system state gets sufficiently close to certain guards.}
\While{$\mathcal{T}\neq\emptyset$}
 \State $d^{\mathcal{T}}\gets\inf\limits_{t\in \mathcal{T}}\inf\limits_{y\in G_{act}^{\ell_i}}\phi^{\ell_i}(\xi^{\ell_i}(t,x_0^i),y)$
 \If{$d^{(k)}\le d^{\mathcal{T}}$}
 \State break the while loop
 \EndIf
\State $k\gets k+1$\Comment{$k$ is the number of pivots.}
\State $t^{(k)}\gets\sup\{\argmin\limits_{t\in\mathcal{T}}\inf\limits_
{y\in G_{act}^{\ell_i}}\phi^{\ell_i}(\xi^{\ell_i}(t,x_0^i),y)\}$
\Comment{At the pivot time instant $t^{(k)}$, the system state gets closest to the guards as $t$ varies in $\mathcal{T}$.} 
\State $G_c^{(k)}\gets ProximalGuards(\ell_i, x_0^i,t^{(k)}, d_i)$
\State 
\State take $\tau_{lead}^{(k)}\in [0,\tau_{maxlead}], \tau_{lag}^{(k)}\in [0,\tau_{maxlag}]$ such that the following conditions are satisfied for all $\tau\in T^{(k)}:=[t^{(k)}-\tau_{lead}^{(k)},t^{(k)}+\tau_{lag}^{(k)}]$ :
\begin{itemize}
\item$G_c^\tau\subset G_c^{(k)}$, where $G_c^\tau\gets ProximalGuards(\ell_i,x_0^i,\tau,d_i)$.
\item$\forall g\in G_c^\tau$, let $(\ell_i,\ell,g,r)$ denote the corresponding event, and $y^{(k)}\gets ProximalState(\ell_i,x_0^i,t^{(k)},g)$, $y^\tau\gets ProximalState(\ell_i, x_0^i,\tau,g)$. Then $\forall g\in G_c^\tau$, it is satisfied that $ y^\tau \in S^{(k)}:= r^{-1}(SafeNeighborhood(\ell,r(y^{(k)}),t^{(k)},t_{end}))$,
and $\phi^{\ell_i}(y^\tau,y^{(k)})\le \alpha\inf\limits_{y\in S^{(k)}}\phi^{\ell_i}(y,y^{(k)})$, where $\alpha\in (0,1)$ is a constant.
\end{itemize}
\State $T^{(k)}\gets T^{(k)}\setminus\bigcup\limits_{j=1}^{k-1}T^{(j)}$
\Comment{$\{T^{(j)}\}_{j=1}^k$ are disjoint.}
\State $\breve{G}^{\ell_i}:=\bigcup\limits_{g\in G_c^{(k)}}g\cap r^{-1}(SafeNeighborhood(\ell,r(y^{(k)}),t^{(k)},t_{end}))$
\Comment{$\forall  g\in G_c^{(k)}, (\ell_i,\ell,g,r)$ is the event; $\check{G}^{\ell_i}$ denotes the allowed part of $G^{\ell_i}$.}
\State $d^{(k)}\gets\inf\limits_{t\in T^{(k)}}\inf\limits_{y\in G_{act}^{\ell_i}\setminus\check{G}^{\ell_i}}\phi^{\ell_i}(\xi^{\ell_i}(t,x_0^i),y)$
\State $\mathcal{T}\gets\mathcal{T}\setminus T^{(k)}$
\EndWhile
\State $\mathring{\Delta}_i:=[t_0^i,t_{end}^i]\setminus\bigcup\limits_{j=1}^k T^{(j)}$, $d_i^g\gets\inf\limits_{t\in \mathring{\Delta}_i}\inf\limits_{y\in G_{act}^{\ell_i}}\phi^{\ell_i}(\xi^{\ell_i}(t,x_0^i),y)$
\State $\gamma_i\gets\min\{d_i^u,d_i^g,d^{(1)},\ldots,d^{(k)}\}$, $\gamma_i\gets Shrinking(\gamma_i)$
\EndFor
\State $\gamma\gets\gamma_1$, $Safe(x_0):=\{x\vert\phi^{\ell_1}(x_0,x)\le\gamma\}$
\State \Return $Safe(x_0)$
\EndProcedure
\end{algorithmic}
\end{algorithm}

\begin{algorithm}
\caption{Subroutine. Obtain guards that are sufficiently close to $\xi^\ell(\tau,x_0)$.}
\label{alg:proxg}
\begin{algorithmic}[1]
\Procedure{ProximalGuards}{$\ell$, $x_0$, $\tau$, $d$}
\State $G_c\gets\{g_{act}\in G_{act}^\ell\vert \inf\limits_{y\in g_{act}}\phi^{\ell}(\xi^{\ell}(\tau,x_0),y)\le d\}$
\State \Return $G_c$ \Comment{Output $G_c$ as the  proximal guards at the time instant $\tau$.}
\EndProcedure
\end{algorithmic}
\end{algorithm}

\begin{algorithm}
\caption{Subroutine. Obtain the state on the guard $g$ that is closest to $\xi^\ell(\tau,x_0)$.}
\label{alg:proxs}
\begin{algorithmic}[1]
\Procedure{ProximalState}{$\ell$, $x_0$, $\tau$, $g$}
\State $Y_c\gets\argmin\limits_{y\in \cl(g)}\phi^{\ell}(\xi^{\ell}(\tau,x_0),y)$
\State $y\gets Y_c$ \Comment{For clarity, we assume $Y_c$ is a singleton. For example, when the guards are hyberplanes, $Y_c$ must be a singleton. If not, the procedure can be extended by choosing a proper $y\in Y_c$.} 
\State \Return $y$ \Comment{Output $y$ as the proximal state at the time instant $\tau$.}
\EndProcedure
\end{algorithmic}
\end{algorithm}

\begin{algorithm}
\caption{Subroutine. Shrink the radius $\gamma_i$ by a proper amount for event time lag compensation \cite{Julius2007}.}
\label{alg:shrinking}
\begin{algorithmic}[1]
\Procedure{Shrinking}{$\gamma_i$}
\State simulate $\xi^{\ell_i}(t,x_0^i)$ for $t_{end}^i\le t\le t_{end}^i+\tau_{maxlag}$ according to the dynamics of location $\ell_i$
\State $\tilde{d}^u_{i}(\tau')\gets\inf\limits_{t\in [t_{end}^i,t_{end}^i+\tau']}\inf\limits_{y\in U^{\ell_i}}\phi^{\ell_i}(\xi^{\ell_i}(t,x_0^i),y)$ for $0\le\tau'\le\tau_{maxlag}$
\State $\mathfrak{T}(\tau'):=[t_{end}^i,t_{end}^i+\tau']\setminus
\bigcup\limits_{j=1}^k T^{(j)}$ \Comment{$\{T^{(j)}\}_{j=1}^k$ are the same as in Algorithm \ref{alg:general}.}
\State $\tilde{d}^g_{i}(\tau')\gets\inf\limits_{t\in \mathfrak{T}(\tau')}\inf\limits_{y\in G_{act}^{\ell_i}}\phi^{\ell_i}(\xi^{\ell_i}(t,x_0^i),y)$ for $0\le\tau'\le\tau_{maxlag}$
\State $\tilde{\gamma}_i(\tau')\gets\min\{\gamma_i,\tilde{d}^u_{i}(\tau'), \tilde{d}^g_{i}(\tau')\}$
\Comment{Clearly, $\tilde{\gamma}_i(0)=\gamma_i$, and $\tilde{\gamma}_i(\tau')$ is non-increasing.} 
\State $d^{inv}_i(\tau')\gets
\sup\limits_{t\in[t_{end}^i,t_{end}^i+\tau']}\inf\limits_{ y\in Inv(\ell_i)}\phi^{\ell_i}(\xi^{\ell_i}(t,x_0^i),y)$ for $0\le\tau'\le\tau_{maxlag}$ \Comment{Clearly, $d^{inv}_i(0)=0$, and $d^{inv}_i(\tau')$ is non-decreasing.}
\State $\mathcal{T}'\gets\{\tau'\in [0,\tau_{maxlag}]
\vert \tilde{\gamma}_i(\tau')\le d^{inv}_i(\tau')\}$
\If{$\mathcal{T}'$ is not empty}
\State  $\tau_{lag}\gets \inf\mathcal{T}'$
\Else
\State $\tau_{lag}\gets\tau_{maxlag}$
\EndIf
\State $\gamma_i\gets d^{inv}_i(\tau_{lag})$
\Comment{$\forall\tau'\in [0,\tau_{lag}]$, $\tilde{\gamma}_i(\tau')\ge d^{inv}_i(\tau')$, which implies $\tilde{\gamma}_i(\tau_{lag})\ge d^{inv}_i(\tau_{lag})=\gamma_i$. So the avoided set cannot be reached before $t_{end}^i+\tau_{lag}$. Besides, $d^{inv}_i(\tau_{lag})=\sup\limits_{t\in [t^i_{end},t^i_{end}+\tau_{lag}]}\inf\limits_{y\in Inv(\ell)}\phi^{\ell_i}(\xi^{\ell_i}(t,x_0^i),y)=\gamma_i$. So any trajectory initiated from the shrunk neighborhood leaves $Inv(\ell)$ before $t_{end}^i+\tau_{lag}$.}
\State \Return $\gamma_i$ \Comment{Output $\gamma_i$ as the radius of the shrunk neighborhood.}
\EndProcedure
\end{algorithmic}
\end{algorithm}
\vspace{-20pt}

\paragraph{Properties of Safe Neighborhoods} The safe neighborhood computed by Algorithm \ref{alg:general} for a general trajectory has the following properties, where Proposition \ref{prop:safe} directly follows from preceding arguments, and Proposition \ref{prop:size} is proved in Appendix.
\begin{proposition}
\label{prop:safe}
For all $\tilde{x}_0\in Safe(x_0)$, the trajectory $\rho(\ell_0,\tilde{x}_0)$ must trigger a path on the event tree that is triggered by the nominal trajectory $\rho(\ell_0,x_0)$ and all its branch trajectories. The time lead/lag for triggering the same event is bounded by $\tau_{maxlead}$ and $\tau_{maxlag}$ respectively. In all locations reached except the last one, $\rho(\ell_0,\tilde{x}_0)$ must stay safe before it leaves the location. In the last reached location, $\rho(\ell_0,\tilde{x}_0)$ must stay safe for at least the same time interval as $\rho(\ell_0,x_0)$ (or its branch trajectory).
\end{proposition}

\begin{definition}[Critical State]
For a guard-critical trajectory, if a state is reached by the trajectory on the closure of guards but does not trigger any event, then it is called a critical state.
\end{definition}

\begin{definition}[Enlarged Reachable Set]
Let $\ell_0$ be an initial location and $Init\subset Inv(\ell_0)$ be a compact initial set of continuous states. 

The enlarged reachable set of an initial state, $Reach^e(x_0)$, is defined as follows:

If the trajectory $\rho(\ell_0,x_0),t_0\le t\le t_{end}$ is not guard-critical, then $Reach^e(x_0)$ only includes the states in $\rho(\ell_0,x_0),t_0\le t\le t_{end}$. Otherwise, $Reach^e(x_0)$ should include the original trajectory as well as all branch trajectories simulated from the critical states for the time horizon $t^*\le t\le t_{end}$, where $t^*$ denotes the time instant when the critical state is reached. 

The enlarged reachable set of an initial set is defined as $Reach^e(Init):=\bigcup\limits_{x_0\in Init}Reach^e(x_0)$.
\end{definition}

\begin{proposition}
\label{prop:size}
The radius of the safe neighborhood computed for $x_0\in Init$ does not vanish if and only if $Reach^e(x_0)\cap\cl(Unsafe)=\emptyset$.
The radii of safe neighborhoods
$\{Safe(x_0)\vert x_0\in Init\}$ are bounded from below by a positive number if and only if
$Reach^e(Init)\cap\cl(Unsafe)=\emptyset$.
\end{proposition}

\subsection{Implementation}
\begin{figure}[H]
\centering
\setlength{\belowcaptionskip}{-10pt}
\includegraphics[scale=0.5]{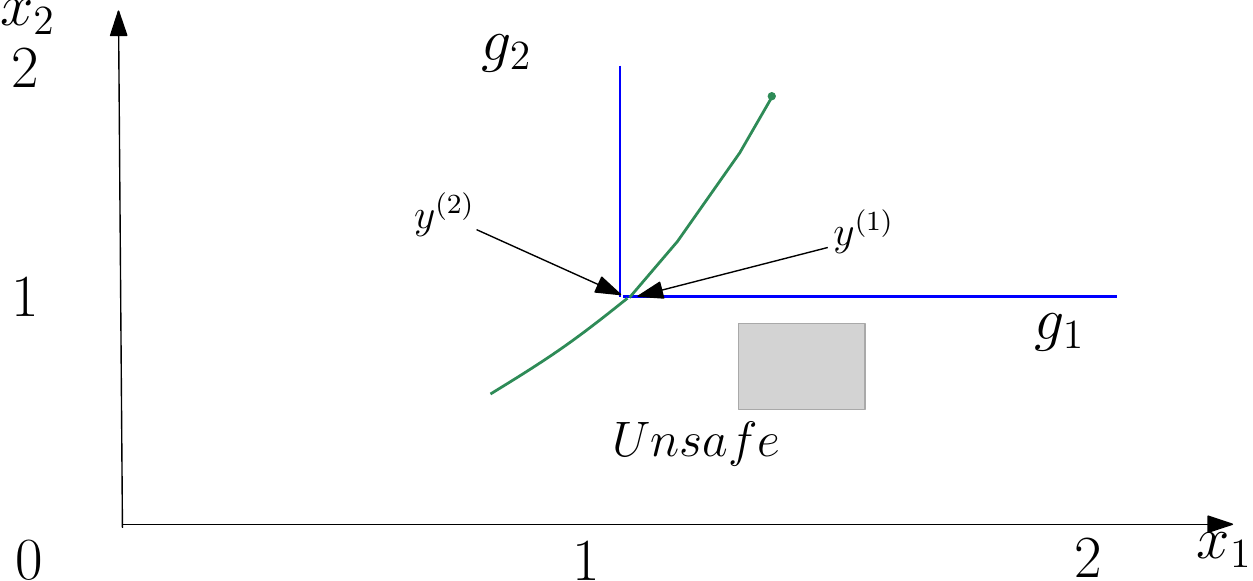}
\caption{A simulated trajectory of the simple example. Locations $\ell_3,\ell_1$ are reached sequentially.} 
\label{fig_simple}
\end{figure}

The robust/safe neighborhood approach is simulation-based, readily parallelizable, and thus suitable for numerical implementation. We have developed a  MATLAB toolbox STRONG (System Testing with RObust Neighborhood Generation)~\cite{Deng2013} that integrates the robust neighborhood and safe neighborhood computation functions for hybrid systems with linear dynamics. 
\vspace{-10pt}
\paragraph{Example}
 In order to illustrate the verification procedure, consider the simple example in Fig. \ref{fig_simple}.
The system has three locations. The invariant sets are $Inv(\ell_1)=Inv(\ell_2)=\mathbb{R}^2$, $Inv(\ell_3)=\{(x_1,x_2)\in\mathbb{R}_2\vert x_1\ge1, x_2\ge 1\}$. Dynamics are $D^{\ell_i}: \dot{x} = A_i x$, where $A_1 = \left(\begin{array}{cc}-1&0\\0&-2\\\end{array}\right), A_2 = \left(\begin{array}{cc}-2&0\\0&-1\\\end{array}\right),A_3 = \left(\begin{array}{cc}-1&0\\0&-3\\\end{array}\right).$ Location $\ell_3$ has guards $g_1 = \{(x_1,x_2)\vert x_1\ge 1,x_2=1\}$ and $g_2 = \{(x_1,x_2)\vert x_1=1,x_2> 1\}$, resetting the discrete state to $\ell_1, \ell_2$ respectively without changing the continuous state. There is an unsafe set $\{\ell_1,\ell_2\}\times\{(x_1,x_2)\vert 1.2\le x_1\le 1.4, 0.5\le x_2\le 0.9\}$. The initial state is $(1.25,1.9)$.
We can simulate a trajectory and compute the robust neighborhood using the command\newline
$>>$ \texttt{traj = RobustTest(sys,sim\_time,max\_lead, max\_lag)},\newline
where \texttt{sys} is the system model, \texttt{sim\_time} is the time horizon $0\le t\le 0.5$, $\texttt{max\_lead}=\texttt{max\_lag}=0.1$ is the maximum event time lead/lag allowed.
The nominal trajectory is shown in Fig. \ref{fig_simple}, for which the radius of robust neighborhood computed as an output of the toolbox is\newline
$>>$ \texttt{traj.ball.d\_min = [0.0042, 0.1613]}.\newline
In the last location reached, $\mathfrak{l}=\ell_1$, there are no guards. The toolbox computes the minimum distance (measured by the bisimulation function $\phi^{\ell_1}$) from the nominal trajectory segment to $Unsafe$, which is $0.1613$. So the robust neighborhood around the reset initial state has radius $0.1613$.

In the initial location $\ell_3$, there are no unsafe states. The toolbox computes the minimum distance (measured by $\phi^{\ell_3}$) to undesired part of guards. The nominal trajectory triggers an event $(\ell_3,\ell_1, g_1, r)$ at $y^{(1)}\in g_1$, where $r$ is identity matrix. Thus, $\check{g}_1:=\{y\in g_1\vert \phi^{\ell_1}(r(y),r(y^*))<0.1613\}$ should be defined as the allowed part of $g_1$. On the other hand, the entire guard $g_2$ is in the avoided set. Since $g_2$ is rather close to the nominal trajectory, the  radius of final robust neighborhood computed around the initial state dramatically shrinks to $0.0042$.

The safe neighborhood computation function is invoked by setting the flag\newline
$>>$ \texttt{sys.opt(1) = true},\newline
and calling the same function \texttt{RobustTest}.

The toolbox will simulate a branch trajectory from $y^{(2)}$ and compute the safe neighborhood around $r(y^{(2)})$, where $r$ is identity matrix. Based on that, part of $g_1$ will be regarded as the allowed part. The bottleneck of minimum distance computation is thus removed. It turns out\newline
$>>$ \texttt{traj.ball.d\_min = [0.0515, 0.1613]},\newline
where $0.0515$ is the radius of final safe neighborhood computed around the initial state.

\section{Conclusion}
The safe neighborhood approach for hybrid automata verification offers mathematically proved guarantee for the safety property of infinitely many initial states by a single trajectory simulation. It inherits the advantages of robust neighborhood approach: no need to grid the state space, and easily parallelizable. The verification procedure has been implemented for linear hybrid systems by the toolbox STRONG.
\bibliographystyle{eptcs}
\bibliography{yd}

\begin{appendices}
\section{Proof of Proposition \ref{prop:size}}
\setcounter{theorem}{8}
\begin{proposition}
The radius of the safe neighborhood computed for $x_0\in Init$ does not vanish if and only if $Reach^e(x_0)\cap\cl(Unsafe)=\emptyset$.
The radii of safe neighborhoods
$\{Safe(x_0)\vert x_0\in Init\}$ are bounded from below by a positive number if and only if
$Reach^e(Init)\cap\cl(Unsafe)=\emptyset$.
\begin{proof}
To prove the first part of the proposition: 

Consider a trajectory with zero radius of safe neighborhood, i.e., $\gamma_1=0$.

According to the subroutine $Shrinking$ in Algorithm \ref{alg:shrinking}, which serves the purpose of event lag compensation, the output $\gamma_1=0$ if and only if the input $\gamma_1=0$. In Algorithm \ref{alg:general}, $\mathcal{T}$ is defined as set of time instants when the system state gets sufficiently close to guards. Clearly for any time instant in $\mathring{\Delta}_1$, the system state is not sufficiently close to guards. Namely, $d^g_1>d_{thr}\ge 0$. 

Suppose the first trajectory segment $\xi^{\ell_1}(t,x^1_0), t_0^1\le t\le t_{end}^1$ does not reach $\cl(Unsafe)$, then $d^u_1>0$. Hence, in location $\ell_1$, $d^{(k)}=0$ for some $k$. There should be some guard $g$ whose closure has zero distance (measured by the bisimulation function $\phi^{\ell_1}$) to the trajectory segment, even if the allowed part has been excluded from the guard. Let $(\ell_1, \ell, g, r)$ denote the corresponding event. In the computation of $d^{(k)}$, $y^{(k)}$ denotes the state on $\cl(g)$ that is closet to the trajectory segment, $t^{(k)}$ denotes the time instant when such a minimum distance is attained, and $S^{(k)}$ denotes the inverse image of the safe neighborhood computed for the reset initial state, i.e, $S^{(k)}:=r^{-1}( SafeNeighborhood(\ell,r(y^{(k)}),t^{(k)},t_{end}))$.
For clarity, we use $d^*, y^*, t^*, S^*$ to replace the notation $d^{(k)}, y^{(k)}, t^{(k)}, S^{(k)}$. 

It follows from $d^*=0$ that $y^*$ is reached by the trajectory segment. So the trajectory simulated from $y^*$ for $t^*\le t\le t_{end}$ (which could be a branch trajectory, or the subsequent segments of the original trajectory) must belong to $Reach^e(x_0)$. Moreover, $d^*=0$ also implies $\inf\limits_{y\in S^*}\phi^{\ell_1}(y,y^*)=0$. By our assumption, the reset map $r$ is a continuous function. It follows that $SafeNeighborhood(\ell,r(y^*),t^*,t_{end})$ must have zero radius.

By preceding arguments, if the safe neighborhood computed for the first segment of the trajectory has zero radius, then either the segment itself reaches $\cl(Unsafe)$, or it reaches the closure of a guard and the safe neighborhood computed around the reset initial state also has zero radius. By induction, if $Safe(x_0)$ has zero radius, there should be a segment of either the original trajectory from $x_0$ or some branch trajectory in $Reach^e(x_0)$ that actually reaches $\cl(Unsafe)$. Therefore, $Safe(x_0)$ is non-trivial as long as $Reach^e(x_0)\cap\cl(U)=\emptyset$. 

It is straightforward that $Reach^e(x_0)\cap\cl(U)\neq\emptyset$ implies trivial $Safe(x_0)$. 
\newline

To prove the second part of the proposition: 

Suppose there exists $\{x_{j}\}_{j=1}^{\infty}\subset Init$ such that $\{\gamma_{j}\}_{j=1}^\infty\rightarrow 0$, where $\gamma_{j}$ denotes the radius of $Safe(x_{j})$. Since $Init$ is compact, there is a subsequence $\{x_j\}_{j=1}^\infty\rightarrow x_0\in Init$ such that $\{\gamma_j\}_{j=1}^\infty\rightarrow 0$ (for brevity, we use the subscript $j$ for all subsequences of $\{x_j\}_{j=1}^\infty$ without changes).

If the radius of a computed safe neighborhood is less than $d_{thr}$, then it must come from $d_{1,j}^u$ or $d^{(k_j)}_j$ for some $k_j$ rather than $d^g_{1,j}$ (the subscript $j$ means the value is corresponding to the initial state $x_j$). For clarity, we use the notation $d^*_j, y^*_j, t^*_j, S^*_j$ to replace such $d^{(k_j)}_j, y^{(k_j)}_j, t^{(k_j)}_j, S^{(k_j)}_j$.

\begin{itemize}
\item 
Suppose as $j$ varies, $d_{1,j}^u$ is bounded from below by a positive number. Since $\{\gamma_j\}_{j=1}^\infty\rightarrow 0$, we can assume without loss of generality that all $\gamma_j$ come from  $d^*_j$ for some $k_j$ instead of $d^u_{1,j}$ or $d^g_{1,j}$.

Since a location has finitely many guards, while there are infinitely many $j$, we can thus assume all $y^*_j$ are on the same guard $g$. $\cl(g)$ is compact, so there is a subsequence  $\{x_j\}_{j=1}^\infty\rightarrow x_0$ such that the corresponding $\{y^*_j\}_{j=1}^\infty$ tends to $y^*_0\in\cl(g)$. 

Clearly, $\{d^*_j\}_{j=1}^{\infty}\rightarrow 0$ implies $\{\inf\limits_{t\in\Delta_j^1}\phi^{\ell_1}(\xi^{\ell_1}(t,x_j), y^*_j)\}_{j=1}^\infty\rightarrow 0$, where $\Delta_j^1$ denotes the dwell time in $\ell_1$ of the trajectory initiated from $x_j$. So $\{\inf\limits_{t\in\Delta_j^1}\phi^{\ell_1}(\xi^{\ell_1}(t,x_j), y^*_0)\}_{j=1}^\infty\rightarrow 0$. It follows from the continuity of the trajectory with respect to the initial condition that
$\inf\limits_{t\in\Delta_0^1}\phi^{\ell_1}(\xi^{\ell_1}(t,x_0), y_0^*)=0$. So the first segment of the trajectory initiated from $x_0$ reaches $y_0^*\in\cl(g)$. The (branch) trajectory simulated from $y_0^*$ must belong to $Reach^e(x_0)$. Let $\gamma^*_j$ denote the radius of $r(S^*_j)$. Following from $\{d^*_j\}_{j=1}^\infty\rightarrow 0$ and the continuity of the reset map $r$, we have $\{\gamma_j^*\}_{j=1}^\infty\rightarrow 0$ and $\{r(y^*_j)\}_{j=1}^\infty\rightarrow r(y^*_0)$. 

\item 
Suppose there exists a subsequence of initial states $\{x_j\}_{j=1}^\infty\rightarrow x_0$ for which $d_{1,j}^u$ tends to $0$. By the continuity of the trajectory with respect to the initial condition we have $\inf\limits_{t\in\Delta_0^1}\inf\limits_{y\in U^{\ell_1}}\phi^{\ell_1}(\xi^{\ell_1}(t,x_0), y)=0$. Namely, the first trajectory segment initiated from $x_0$ reaches $\cl(Unsafe)$.
\end{itemize}

By preceding arguments,  if $\{x_j\}_{j=1}^\infty\rightarrow x_0, \{\gamma_j\}_{j=1}^\infty\rightarrow 0$, then either the first segment of the trajectory initiated from $x_0$ reaches $\cl(Unsafe)$, or there exist $\{r(y^*_j)\}_{j=1}^\infty\rightarrow r(y^*_0), \{\gamma^*_j\}_{j=1}^\infty\rightarrow 0$ such that the trajectory simulated from $y^*_0$ belongs to $Reach^e(x_0)$. Using induction, it can be proved $\{x_j\}_{j=1}^\infty\rightarrow x_0, \{\gamma_j\}_{j=1}^\infty\rightarrow 0$ implies there must be some trajectory segment in $Reach^e(x_0)$ that actually reaches $\cl(Unsafe)$. Therefore, the radii of safe neighborhoods $\{Safe(x_0)\vert x_0\in Init\}$ are bounded from below by a positive number as long as $Reach^e(Init)\cap\cl(Unsafe)=\emptyset$.

The converse direction is straightforward.
\end{proof}
\end{proposition}
\end{appendices}
\end{document}